\documentclass[a4paper,10pt,aps,prd,twocolumn]{revtex4}
\usepackage[colorlinks,linkcolor={red},citecolor={blue}]{hyperref}
\usepackage{graphicx}
\usepackage{amsmath}
\usepackage{amssymb}
\usepackage{amsfonts}
\usepackage{bm}
\usepackage{braket}
\usepackage{color}

\def\be{\begin{align}}
\def\ee{\end{align}}
\def\bea{\begin{eqnarray}}
\def\eea{\end{eqnarray}}

\def\e{\mathrm{e}}

\newcommand{\mc}[1]{\mathcal{#1}}
\newcommand{\f}[2]{\frac{#1}{#2}}
\begin{document}
	
\title{Cosmic acceleration via derivative matter couplings}
\author{Zahra Haghani}
\email{z.haghani@du.ac.ir}
\author{Shahab Shahidi}
\email{s.shahidi@du.ac.ir}
\affiliation{School of Physics, Damghan University, Damghan, Iran.}

\begin{abstract}
	A gravitational theory with derivative matter coupling is proposed which adopt de Sitter expansion at late times with ordinary baryonic matter. Matter components are conserved separately in the cosmological background and the Newtonian limit is well behaved. Also Conformal transformation of the theory to Jordan frame showing an effective cosmological constant explaining the nature of de Sitter acceleration.
\end{abstract}

\pacs{}

\maketitle
\section{Introduction}
One of the earliest attempts to modify the Einstein's gravitational theory was done by Einstein himself \cite{Einstein} where he tried to solve the problem of the structure of the elementary particles. He assumed that the elementary particles behave like radiation and contribute to the Einstein's equations through the Maxwell's energy momentum tensor $T^e_{\mu\nu}$. Since the energy momentum tensor of the Maxwell field is traceless, he modified the geometry part of his gravitational equations to become traceless and write the dynamical equations as
$$R_{\mu\nu}+\bar\lambda g_{\mu\nu}R=T^e_{\mu\nu},$$
where $\bar{\lambda}=-1/4$. This formalism straightforwardly explains the cosmological constant as an integration constant \cite{PDU}. The idea was generalized to an arbitrary matter field with energy momentum tensor $T_{\mu\nu}$ (not necessarily radiation) an arbitrary $\bar{\lambda}$‌ by Rastall \cite{rastall}. Obviously in Rastall's theory the matter energy momentum tensor is not  conserved and as a result there is a possibility of creation of matter from the gravitational field. Rastall's theory is a special case of a class of theories which is now known as $f(R,T)$ theories, where $T$‌ is the trace of the energy momentum tensor \cite{fRT}. This is however one of the classes of theories representing the matter-geometry couplings. Other classes could contain an arbitrary function of the matter Lagrangian $L_m$ \cite{Lmtheory} or contain more complicated matter-geometry couplings like $R_{\mu\nu}T^{\mu\nu}$ where $R_{\mu\nu}$‌ is the Ricci tensor \cite{ourRT}. Also one can consider higher order matter couplings with Lagrangian density $\sqrt{-g}(R+f(T_{\mu\nu}T^{\mu\nu}))$ \cite{roshan,powered}. It is obvious that the non-minimal matter-geometry coupling can in principle produce dS ‌expansion. Recently however, it was suggested that a minimal coupling between geometry and matter with Lagrangian $\sqrt{-g}(R+(T_{\mu\nu}T^{\mu\nu})^\eta)$ can produce accelerated expansion which is in agreement with recent observational data \cite{powered}. However, these minimal geometry-matter coupling can not produce de Sitter (dS) expansion as we will discuss in the next section. In this paper, we will propose a model with minimal geometry-matter couplings with the property that in the late time the universe experiences a dS expansion. The cost we should pay is to enforce the matter sector to contain derivative matter couplings. It should be noted that in all of the above classes of theories, because we have a matter action $S_m$ added linearly to the theory, we can not find an exact de Sitter solution in the presence of matter. However, at late times where the energy density tends to zero, the effect of $S_m$‌ can be ignored and one could have a dS solution.
\section{The model}
Consider a generic gravitational action with matter, consists of an arbitrary function of ordinary matters
\begin{align}
S=\int d^4x\sqrt{-g}\bigg[\kappa^2 R+f(T,T_{\mu\nu}T^{\mu\nu},L_m)\bigg]+S_m,
\end{align}
where $f$‌ does not contain derivatives. The Friedmann and Raychaudhuri equations can be written generically as
\begin{align}
3H^2&=\f{1}{2\kappa^2}\rho+g_1(\rho,p),\nonumber\\
-2\dot{H}-3H^2&=\f{1}{2\kappa^2}(\rho+3p)+g_2(\rho,p),
\end{align}
where $g_i$‌ are some functions of the energy density and pressure. We are seeking the dS expansion at late times from the above equations. This means that the functions $g_i$‌ should becomes constant at $a\rightarrow\infty$. Assuming that the matter contents of the universe are ordinary baryonic matter with $\rho,p\rightarrow0$ at late times, one can deduce that the above equations will give late-time de Sitter expansion only if $g_i=const.$ which implies $f=const.$ In summary, an arbitrary algebraic minimal matter extension of the Einstein-Hilbert theory can not produce dS expansion. Note that using non-minimally coupling between matter and geometry $f(R,T,...)$ the dS expansion can be achieved. 

In this paper however, we want to obtain dS expansion using minimal coupling between matter and geometry. As discussed above, one should add to the action some non-algebraic function of the matter content. Let's try some derivative coupling of the matter Lagrangian $L_m$: $f(L_m,\nabla L_m)$. Also assume that the matter Lagrangian is $L_m=-\rho$ corresponding to the perfect fluid and  the energy density behaves as $\rho\propto a^{-n}$ ($n=3,4$ for dust and radiation respectively). In order to have a dS expansion, $f$ should become constant as discussed above. Because the matter Lagrangian is a scalar, $f‌$ should contain even number of derivatives. The simplest term could be $\nabla_\mu\rho\nabla^\mu\rho$ which behaves like $H^2\rho^2$ and vanishes at late times. The main problem with this term is the appearance of $\rho^2$ which can be solved if one uses $\ln\rho$ instead of $\rho$. As a result $\nabla_\mu\ln\rho\nabla^\mu\ln\rho$ could behaves as an effective cosmological constant at late times and produces a dS expansion.

With the aid of above discussions, let us consider the following action
\begin{align}\label{action}
S=\int d^4x\sqrt{-g}\bigg[\kappa^2 R&+\alpha\,\nabla_\mu f\,\nabla^\mu f\nonumber\\&+\beta \,\Box f \,\nabla_\mu f\,\nabla^\mu f\bigg]+S_m,
\end{align}
where we have defined $f=\Lambda\ln(|L_m|/\Lambda^4)$. The second term in the above action could be considered as a canonical kinetic term of the scalar function $f$ and the third term is the cubic Galileon term which is a ghost free higher derivative self interaction of $f$ and we have added to the action for completeness. Note that $\Lambda$ is a constant with mass dimension $M$ in order to make the argument of the Logarithm dimensionless. We will see in the next section that the value of $\Lambda$ should be of order of $\kappa$. Also $\beta$ should have a mass dimension $M^{-3}$ and $\alpha$ is dimensionless.

In order to find the equations of motion of the above action, we first note that from the definition of the energy momentum tensor
\begin{align}
T_{\mu\nu}=\f{-2}{\sqrt{-g}}\f{\delta(\sqrt{-g}L_m)}{\delta g^{\mu\nu}},
\end{align}
one can obtain the variation of the matter Lagrangian as
\begin{align}
	\delta L_m=\f{1}{2}\big(L_m g_{\mu\nu}-T_{\mu\nu}\big)\delta g^{\mu\nu}.
\end{align}
With this in hand, one can obtain the equation of motion of the metric field as
\begin{widetext}
\begin{align}\label{eom}
\kappa^2G_{\mu\nu}&-\f12T_{\mu\nu}+\alpha\left(\nabla_\mu f\nabla_\nu f-\f12g_{\mu\nu}\nabla_\alpha f\nabla^\alpha f \right)+\beta\bigg(g_{\mu\nu}\nabla^\alpha f\nabla_\alpha\nabla_\beta f\nabla^\beta f+\Box f \nabla_\mu f \nabla_\nu f-2\nabla^\alpha f\nabla_\alpha\nabla_{(\mu}f\nabla_{\nu)}f \bigg)\nonumber\\&+\f{\alpha\Lambda}{L_m}(T_{\mu\nu}-g_{\mu\nu}L_m)\Box f+\f{\beta\Lambda}{L_m}(T_{\mu\nu}-g_{\mu\nu}L_m)\bigg((\Box f)^2-\nabla_\alpha\nabla_\beta f\nabla^\alpha\nabla^\beta f-R_{\alpha\beta}\nabla^\alpha f\nabla^\beta f\bigg)=0.
\end{align}
\end{widetext}
The conservation of the energy momentum tensor can be obtained as
\begin{align}\label{conser}
\nabla^\mu &T_{\mu\nu}=\f{2(L_mg_{\mu\nu}-T_{\mu\nu})}{1-\alpha\Lambda\Box f-\beta\Lambda\Omega}\nonumber\\&\times\Big[\alpha(\nabla^\mu f\Box f-\Lambda\nabla^\mu\Box f)+\beta(\Omega\nabla^\mu f-\Lambda\nabla^\mu\Omega)\Big],
\end{align}
where we have defined
\begin{align}
	\Omega=(\Box f)^2-\nabla_\alpha\nabla_\beta f\nabla^\alpha\nabla^\beta f-R_{\alpha\beta}\nabla^\alpha f\nabla^\beta f
\end{align}
One should note here that despite the fact that the energy momentum tensor is not conserved in general, it behaves differently in the cosmological setup. In fact, assuming a perfect fluid with energy density $\rho$‌ and pressure $p$, one can easily verify that \eqref{conser} implies
$$\dot{\rho}+3H(\rho+p)=0,$$
stating that in the cosmological setup, the energy momentum tensor of the perfect fluid is conserved.
\section{Newtonian limit}
Let us consider the Newtonian limit of the theory \eqref{action}. After computing the trace of equation \eqref{eom} and substituting the Ricci scalar back to \eqref{eom}, one obtains an equivalent form of the equations of motion as
\begin{align}\label{eomtr}
\kappa^2& R_{\mu\nu}-\f12\left(T_{\mu\nu}-\f12g_{\mu\nu}T\right)+\alpha(\Lambda g_{\mu\nu}\Box f+\nabla_\mu f\nabla_\nu f)\nonumber\\
&-\f{\beta\Lambda}{2L_m}(g_{\mu\nu}T-2T_{\mu\nu}-2g_{\mu\nu}L_m)\Big((\Box f)^2\nonumber\\&-R_{\alpha\beta}\nabla^\alpha\nabla^\beta-\nabla_\alpha\nabla_\beta f\nabla^\alpha\nabla^\beta f\Big)+\beta\Big(\Box f\nabla_\mu f\nabla_\nu f\nonumber\\
&-2\nabla_\alpha f\nabla_\alpha\nabla_{(\mu)}f\nabla_{\nu)}-\f12g_{\mu\nu}\Box f\nabla_\alpha f\nabla^\alpha f\Big)\nonumber\\&-\f{\alpha\Lambda}{2L_m}(g_{\mu\nu}T-2T_{\mu\nu})\Box f=0.
\end{align}
In the Newtonian limit, the gravitational field is weak and the velocity of the particles are small compare to the speed of light. In this regard, the metric can be written as
\begin{align}
ds^2=-(1+2\Phi(\vec{x}))dt^2+d\vec{x}^2,
\end{align}
and the only non-zero component of the energy momentum tensor is $T_{00}=\rho$. Plugging the above ansatz into the equation of motion \eqref{eomtr} and keeping up to first order in $\rho$‌ and $\Phi$, one obtains the Poison's equation as
\begin{align}\label{new}
\vec{\nabla}^2\Phi=\f{1}{4\kappa^2}\rho-\f{\rho_0}{\kappa^2}\Phi+\f{3\alpha\Lambda^2}{2\kappa^2}\f{\vec{\nabla}^2\rho}{\rho_0},
\end{align}
where $\rho_0$ is the background value for the energy density corresponding to the flat space. As a result, the value of $\rho_0$‌ should vanish. Tending $\rho_0\rightarrow0$, one can see that the second term of the RHS becomes negligible. In this limit however, the last term in the RHS of equation \eqref{new} behaves like an effective cosmological constant. In this paper however, we are interested to the cosmological applications of the theory. Assuming that the energy density $\rho$‌ vary slowly, one can estimate the cosmological constant as  $\Lambda_{eff}\sim\alpha\Lambda^2 H_0^2/\kappa^2$ where $H^{-1}_0$ is the Hubble horizon. Assuming $\alpha$ to be of order of unity and noting that the value of the cosmological constant is of order of the Hubble horizon, one deduces that $\Lambda\sim \kappa$‌ to have a $\Lambda$CDM expansion at late times.
\section{Cosmological implications}
let us consider the cosmological implications of the model \eqref{action}. Suppose that the universe can be described by the flat FRW ansatz
\begin{align}
ds^2=-dt^2+a^2(dx^2+dy^2+dz^2),
\end{align}
filled with a perfect fluid with matter Lagrangian of the form $L_m=-\rho$ and the energy-momentum tensor
\begin{align}
T^\mu_\nu=\textmd{diag}(-\rho,p,p,p),
\end{align} 
where $\rho=\rho(t)$‌ is the matter energy density and $p=p(t)$‌ is the pressure. Subsitituting in \eqref{eom}, one can obtain the Friedman and Raychaudhuri equations as
\begin{align}\label{frid}
&\rho+ 2 \alpha  \Lambda   \ddot{f} -\dot{f}^2
\left(\alpha +6 \beta  \Lambda   \dot{H} \right)-6 H^2 \left(3 \beta  \Lambda 
\dot{f}^2+\kappa ^2\right)\nonumber\\&+6 H   \dot{f}  \left(\Lambda  \left(\alpha -2 \beta 
\ddot{f} \right)+\beta   \dot{f}^2\right)  +\frac{1}{ L_m }\bigg[-18 \beta  \Lambda  H^2 \rho    \dot{f}^2 \nonumber\\&+2 \Lambda  \rho   \left(\alpha 
\ddot{f}-3 \beta   \dot{f}^2  \dot{H} \right)+6 \Lambda  H  \rho    \dot{f}  \left(\alpha -2
\beta   \ddot{f} \right)\bigg]=0,
\end{align}
and
\begin{align}\label{ray}
& \alpha  \Lambda   \ddot{f} -  \dot{H}  \left(3 \beta  \Lambda   \dot{f}^2+2 \kappa^2\right)-3
H^2 \left(3 \beta  \Lambda   \dot{f}^2+\kappa^2\right)+\f{1}{2}\alpha   \dot{f}^2\nonumber\\&+\frac{1}{L_m }\bigg[6 \beta
\Lambda  H^2 p   \dot{f}^2+ \Lambda  p  \left(3 \beta   \dot{f}^2  \dot{H} -\alpha 
\ddot{f} \right)\nonumber\\&-3 \Lambda  H  p   \dot{f}  \left(\alpha -2 \beta 
\ddot{f} \right)\bigg]+3 \Lambda  H   \dot{f}  \left(\alpha -2 \beta   \ddot{f} \right)\nonumber\\&+
\beta   \dot{f}^2  \ddot{f} -\f12 p =0,
\end{align}
where dot denotes time derivative. In order to close the above system, we will also assume that the perfect fluid obeys the barotropic equation of state, $p=\omega\rho$. Let us separately investigate the possibility of having late time dS expansion in the dust/radiation dominated universe. We also use the conservation of the energy momentum tensor in the cosmological setup as discussed in the last section.
\subsection{Dust dominated universe}
In the dust dominated universe with $\omega=0$, the conservation of the energy momentum tensor implies $\rho=\rho_0 /a^3$ where $\rho_0$ is some positive constant. Equations \eqref{frid} and \eqref{ray} becomes
\begin{align}\label{d1}
\frac{\rho _0}{2 a^3}-\frac{9}{2} \alpha  H_0^2 \Lambda ^2-81 \beta  H_0^4 \Lambda
^3-3 H_0^2 \kappa ^2=0,
\end{align}
and
\begin{align}\label{d2}
-\frac{9}{2} \alpha  H_0^2 \Lambda ^2-81 \beta  H_0^4 \Lambda ^3-3 H_0^2 \kappa ^2=0,
\end{align}
where we have set $H=H_0=const.$
At late times the first term in \eqref{d1} can be neglected and the above equations has a solution
\begin{align}\label{hdust}
H_0^2=-\frac{3 \alpha  \Lambda ^2+2 \kappa ^2}{54 \beta  \Lambda ^3}.
\end{align}
Assuming the estimated value for $\Lambda\sim\kappa$ from the Newtonian limit considerations, one can easily see that at late times $\beta\kappa\sim H_0^{-2}$ with $H_0=68.7\pm4.5\, \textmd{km} \,\textmd{s}^{-1} \textmd{Mpc}^{-1}$ .

For $\alpha=-1/2$ which makes the kinetic term in \eqref{action} canonical, we obtain the constraint
$0<\Lambda<\f{2}{\sqrt{3}}\kappa$ for $\beta<0$ and $\Lambda>\f{2}{\sqrt{3}}\kappa$ for $\beta>0$ in order to make the Hubble parameter real. The above discussions then shows that negative values of $\beta$ is more favored from cosmological point of view.
 It should be noted that the presence of the Galileon term in \eqref{action}, does not change the result. In fact, for $\beta=0$ one has $\alpha= -2 \kappa ^2/3 \Lambda ^2$ with $H_0$ remains arbitrary and can be set from observational data to be $H_0=68.7\pm4.5\, \textmd{km} \,\textmd{s}^{-1} \textmd{Mpc}^{-1}$. This gives $\alpha$ to be negative and of order of unity for $\Lambda\sim\kappa$ in agreement with the Newtonian limit considerations. 
 
To see the stability of this solution let us perturb the energy density and Hubble parameter as
\begin{align}
\rho=\f{\rho_0}{a^3}+\delta \rho, \quad H=H_0+\delta H.
\end{align}
Applying the above perturbations into field equations and expanding them to the first order in perturbations, one can obtain the equations at late time limit as

\begin{align}
3 H_0 \Lambda ^2 \left(\delta \dot{\rho }+3 H_0 \delta \rho  \right) \left(\alpha+27
\beta \Lambda H_0^2  \right)=0,
\end{align}
with the solution
\begin{align}
\delta \rho= \delta\rho_0  \ \e^{-3H_0 t},
\end{align}
where $\delta\rho_0$ is constant.This solution shows that the dS solutions is stable at late times.

\subsection{Radiation dominated universe}
Now consider the radiation dominated universe with $\omega=1/3$ with the density profile $\rho=\rho_0/a^4$. The equations of motion \eqref{frid} and \eqref{ray} reduces to
\begin{align}
\frac{\rho _0}{2 a^4}-8 \alpha  H_0^2 \Lambda ^2-192 \beta  H_0^4 \Lambda ^3-3 H_0^2
\kappa ^2=0,
\end{align}
and
\begin{align}
\frac{\rho _0}{6 a^4}+8 \alpha  H_0^2 \Lambda ^2+192 \beta  H_0^4 \Lambda ^3+3 H_0^2
\kappa ^2=0.
\end{align}
Neglecting the first term, one obtains
\begin{align}\label{hrad}
H_0^2= -\frac{8 \alpha  \Lambda ^2+3 \kappa ^2}{192 \beta  \Lambda ^3}.
\end{align}
In the case $\alpha=-1/2$, one obtains the $0<\Lambda<\f{\sqrt{3}}{2}\kappa$ for $\beta<0$ and 
$\Lambda>\f{\sqrt{3}}{2}\kappa$ for $\beta>0$. As in the matter dominated case the Newtonian limit implies that $\beta$‌ should be negative. In the case $\beta=0$, one obtains $\alpha=-3\kappa^2/8\Lambda^2$. 

Upon perturbing the energy density and pressure as
\begin{align}
\rho=\f{\rho_0}{a^4}+\delta \rho, \quad H=H_0+\delta H,
\end{align}
one obtains the late time limit equations of the density perturbations as
\begin{align}
4 H_0 \Lambda ^2 \left(\delta \dot{\rho}+4  H_0 \delta \rho \right) \left(\alpha+36
\beta \Lambda H_0^2  \right)=0,
\end{align}
with a stable solution
\begin{align}
\delta \rho= \delta\rho_0  \ \e^{-4 H_0 t},
\end{align}
where $\delta\rho_0$ is constant.

One can see that the qualitative behavior of the radiation dominated universe is the same as of dust dominated universe. 
\subsection{Numerical results}
Leaving the energy density and the Hubble parameter arbitrary, the equations of motion can be obtained as
\begin{align}
-\frac{\alpha  \Lambda ^2 \dot{\rho }^2}{2 \rho ^2}-3 H^2 \kappa ^2+\frac{3 \beta  H
	\Lambda ^3 \dot{\rho }^3}{\rho ^3}+\frac{\rho }{2}=0,
\end{align}
and
\begin{align}
&\frac{1}{2 \rho ^4}\bigg[-\Lambda ^2 \rho  \dot{\rho }^2 \left(\rho  \left(2 \alpha  \omega +\alpha +18
\beta  H^2 \Lambda  (\omega +1)\right)-2 \beta  \Lambda  \overset{\text{..}}{\rho
}\right)\nonumber\\&+\rho ^3 \left(2 \alpha  \Lambda ^2 (\omega +1) \overset{\text{..}}{\rho
}-\rho  \left(6 H^2 \kappa ^2+\rho  \omega \right)\right)-2 \beta  \Lambda ^3 \dot{\rho }^4\nonumber\\&+6 H \Lambda ^2 \rho ^2
\dot{\rho } (\omega +1) \left(\alpha  \rho -2 \beta  \Lambda  \overset{\text{..}}{\rho
}\right)+12 \beta  H \Lambda ^3 \rho 
\dot{\rho }^3 (\omega +1)\nonumber\\&-2 \dot{H} \rho ^2 \left(3 \beta  \Lambda ^3
\dot{\rho }^2 (\omega +1)+2 \kappa ^2 \rho ^2\right)\bigg]=0.
\end{align}
Defining the following set of dimensionless parameters
\begin{align}
&H=H_{0}h,\quad\rho=6\kappa^2 H_0^2 r,\qquad \tau=H_0 t,\nonumber\\
&\beta=\f{\gamma}{\kappa H_0^2},\quad \Lambda=\kappa \lambda,
\end{align}
one can write the dimensionless equations of motion as
\begin{align}
-3 h^2+3 r+\frac{3 h \epsilon  r'^3}{r^3}-\frac{\eta  r'^2}{2
	r^2}=0,
\end{align}
and
\begin{align}
&\frac{1}{2 r^4}\bigg[6 h^2 r^2 \left(r^2+3 (\omega +1) \epsilon  r'^2\right)+4 r^4 h'\nonumber\\&-6 h r
(\omega +1) r' \left(\eta  r^2-2 r \epsilon  r''+2 \epsilon 
r'^2\right)+6 r^5 \omega +2 \epsilon  r'^4\nonumber\\&+r^2
r'^2 \left(2 \eta  \omega +\eta +6 (\omega +1) \epsilon  h'\right)-2 \eta  r^3
(\omega +1) r''\nonumber\\&\qquad\qquad-2 r \epsilon  r'^2 r''\bigg]=0,
\end{align}
where prime denotes derivative with respect to $\tau$ and we have defined  
$\eta= \lambda^2 \alpha$ and $\epsilon=\lambda \gamma^3$.

\begin{figure}\label{hub1}
	\includegraphics[scale=0.4]{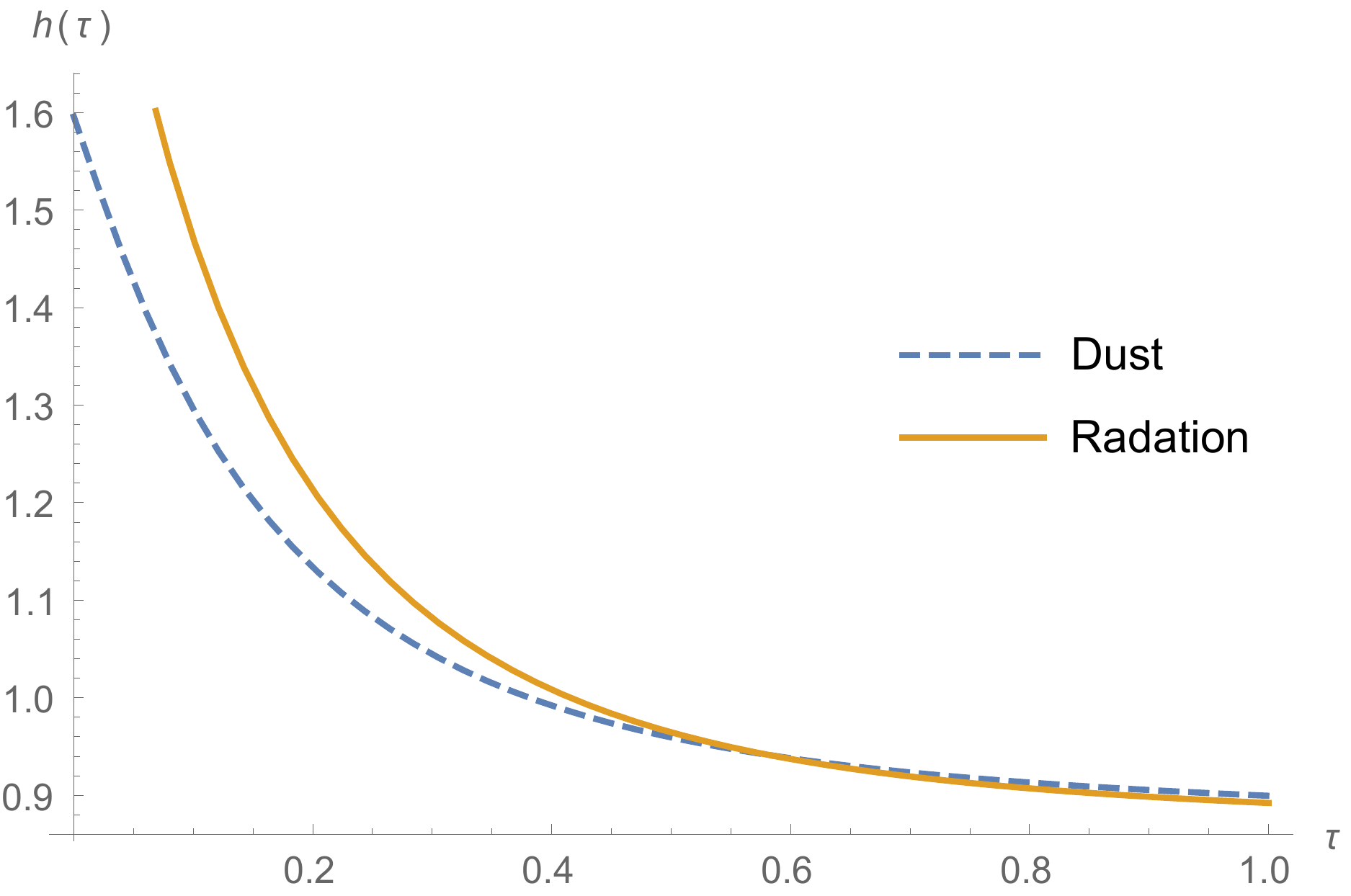}
	\includegraphics[scale=0.4]{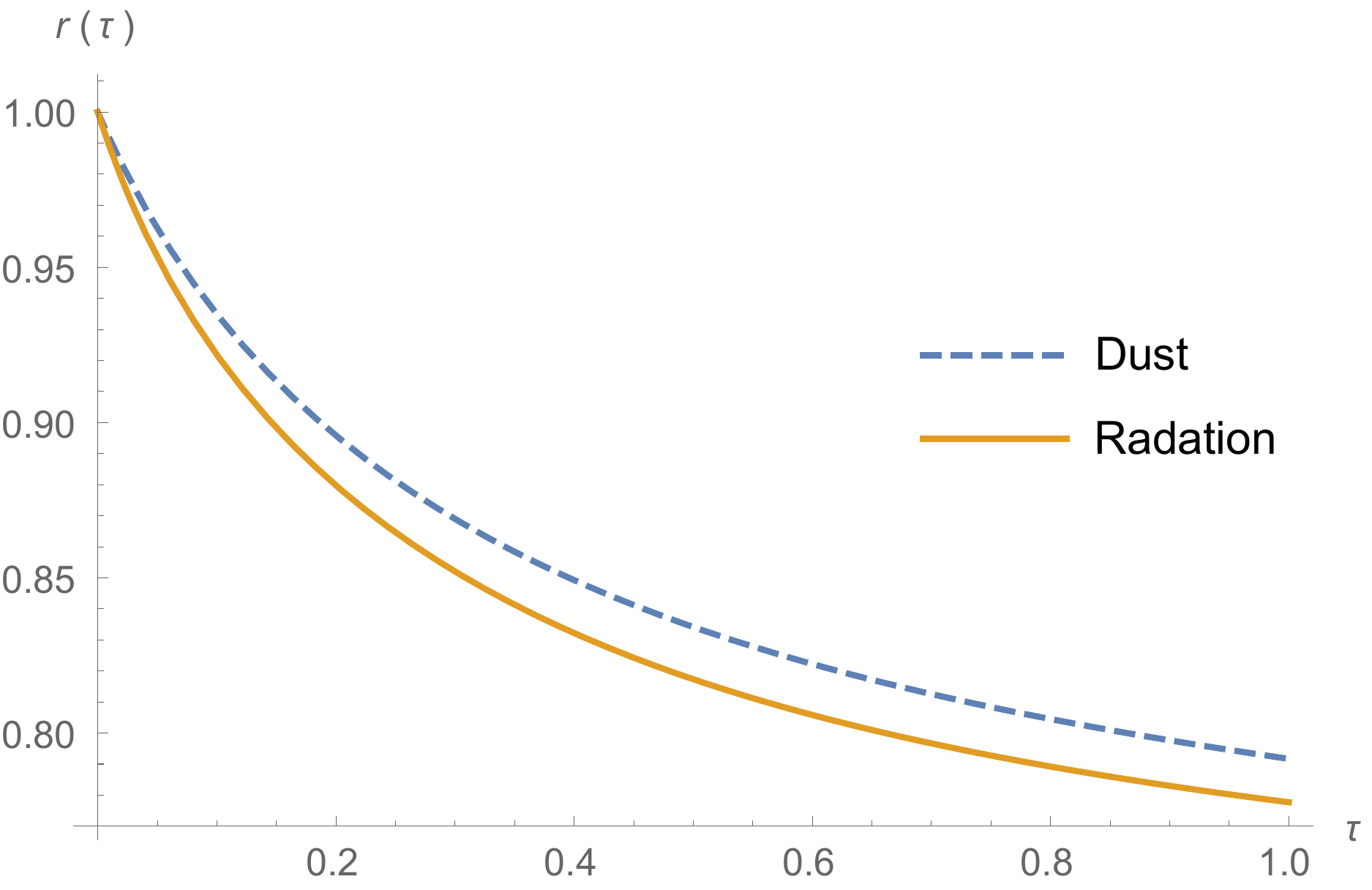}
	\caption{Plot of the dimensionless Hubble parameter $h$ and the energy density $r$ for dust/radiation dominated universe. We have assumed  $\epsilon=1.150$ and $\eta=-21.3667 (-27.975)$ for dust (radiation) dominated universe respectively.}
\end{figure}

\begin{figure}\label{dec}
	\includegraphics[scale=0.4]{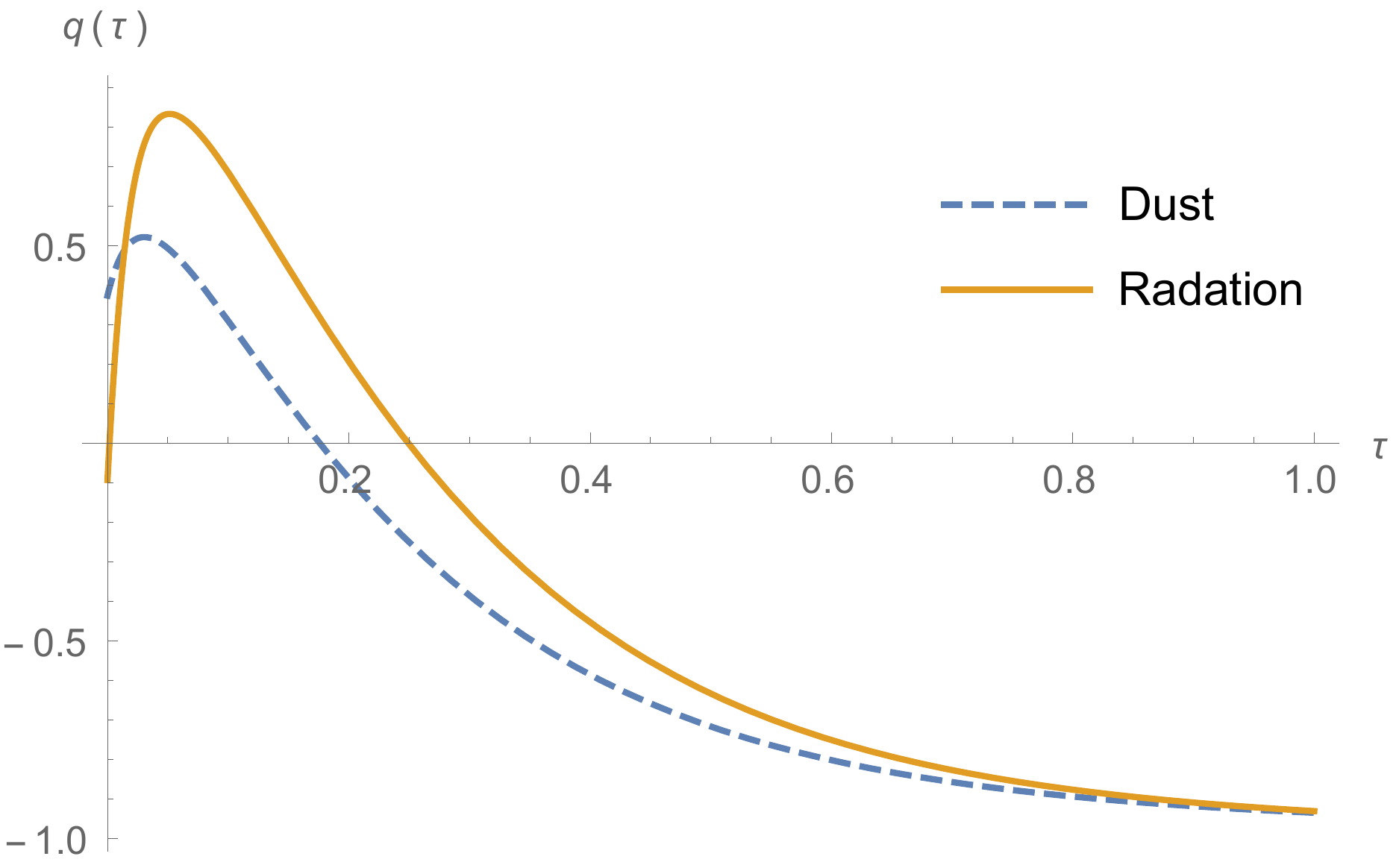}
	\caption{Plot of the deceleration parameter $q$ for dust/radiation dominated universe. We have assumed  $\epsilon=1.150$ and $\eta=-21.3667 (-27.975)$ for dust (radiation) dominated universe respectively.}
\end{figure}

Figures \eqref{hub1} and \eqref{dec} shows the behavior of the Hubble parameter, the energy density and the deceleration parameter $q=-1-\dot{H}/H^2$ for dust and radiation dominated universe. In the figures we have used $\epsilon=1.150$ and find the initial value of $\eta$‌ from the Friedman equation which is $\eta=-21.3667$ for dust dominated universe and $\eta=-27.975$ for radiation dominated universe. One should note that the value of $\beta$‌ is assumed to be negative in agreement with the discussions of the previous section. It is evident from the figures that the late time behavior of the universe is consistent with the dS expanding phase as we have argued in this paper.

\section{Dynamical system analysis}
\begin{figure}\label{figdyna}
	\includegraphics[scale=0.65]{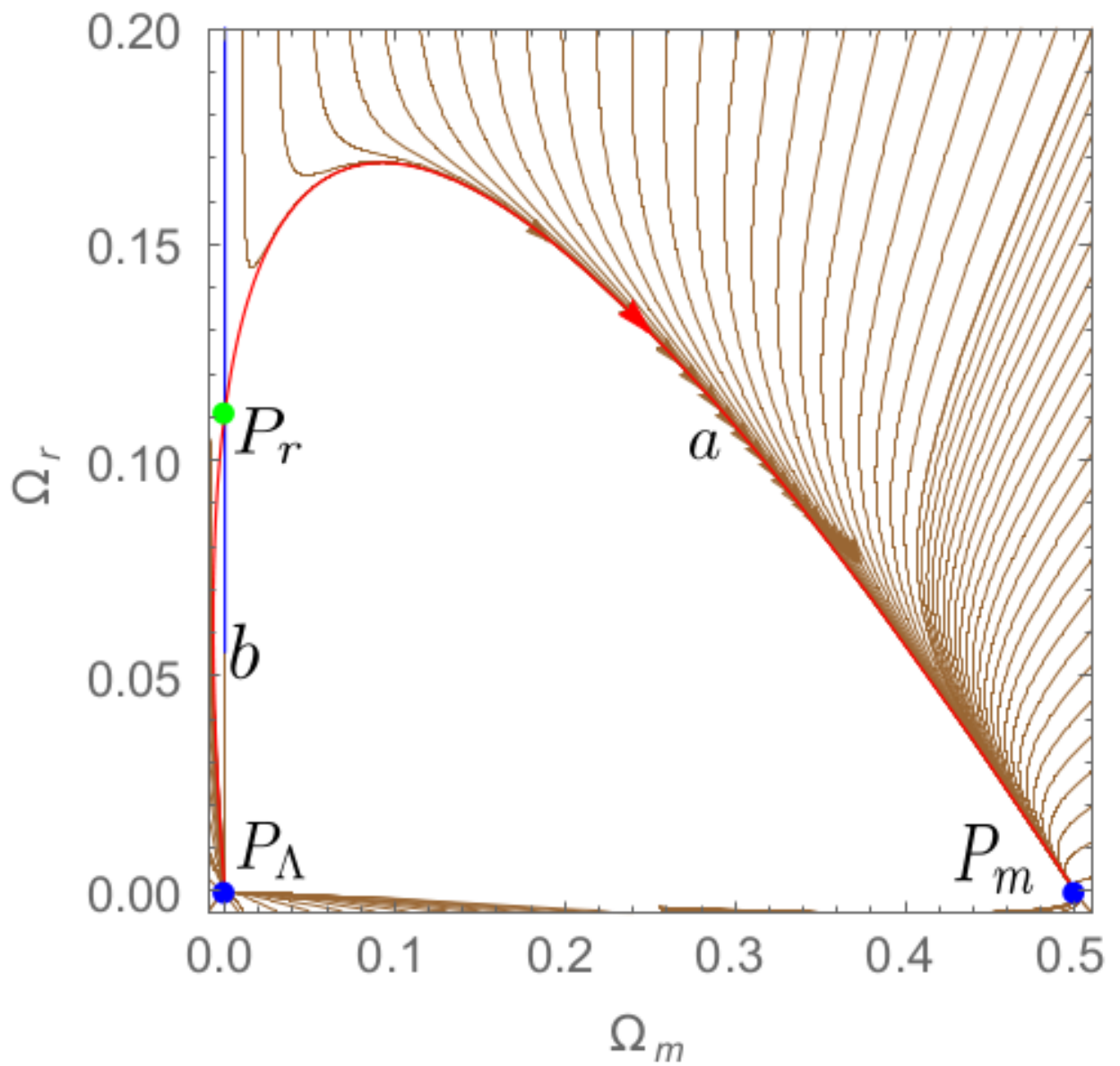}
	\caption{The phase portrait corresponds to the dynamical system \eqref{dy1} and \eqref{dy2}. Three fixed points are denoted by $P_r$, $P_m$ and $P_\Lambda$. The blue/red lines represents the stable/unstable manifolds of the fixed point $P_r$ respectively.}
\end{figure}

Assume that the universe filled with dust and radiation
$\rho=\rho_r+\rho_m,$
where $\rho_r$ and $\rho_m$ are the energy density of the relativistic and non-relativistic matter respectively.
As was pointed out in the previous section, these two components are conserved on top of the cosmological background
\begin{align}
\dot{\rho}_i + 3(\omega_i +1) H \rho_i=0, \qquad i=r,m
\end{align}
where the equation of state parameters are $\omega_r=1/3$ and $\omega_m =0$. Now define dynamical variables
\begin{align}\label{var1}
\Omega_i=\frac{\rho_i}{6 \kappa ^2 H^2},\quad \Omega_H=\frac{\kappa ^2}{H^2},
\end{align}
with $i=r,m$.
Equation \eqref{frid} shows that $\Omega_H$ is not independent
\begin{align}
\Omega_H=&6 \beta _1 \left(3 \Omega _m+4 \Omega _r\right){}^3\bigg(6 \left(\Omega _m+\Omega
_r-1\right) \left(\Omega _m+\Omega _r\right){}^3\nonumber\\&-\alpha _1 \left(\Omega _m+\Omega
_r\right) \left(3 \Omega _m+4 \Omega _r\right){}^2\bigg)^{-1},
\end{align}
where we have defined $\beta_1=\beta \Lambda^3$ and $\alpha_1=\alpha\Lambda^2/\kappa^2$.
As a result we obtain a 2-dimensional dynamical system of the form
\begin{align}\label{dy1}
\Omega^{\prime}_m =\f{\Omega _m}{A}&\bigg[ \alpha _1 \left(3 \Omega _m+4 \Omega _r\right){}^2 \left(22 \Omega _m
\Omega _r+9 \Omega _m^2+12 \Omega _r^2\right)\nonumber\\-&6 \left(\Omega _m+\Omega _r\right){}^2
\big(\left(29 \Omega _m-12\right) \Omega _r^2+8 \Omega _r^3 \nonumber\\+&9 \left(\Omega _m-1\right) \Omega _m^2+6 \Omega _m \left(5 \Omega _m-4\right)
\Omega _r \big)\bigg],
\end{align}
\begin{align}\label{dy2}
\Omega^{\prime}_r =\f{\Omega _r}{A}& \bigg[\alpha _1 \left(3 \Omega _m+4 \Omega _r\right){}^2 \left(29 \Omega _m
\Omega _r+12 \Omega _m^2+16 \Omega _r^2\right)\nonumber\\-&6 \left(\Omega _m+\Omega _r\right){}^2
\big(2 \Omega _m^2 \left(25 \Omega _r-6\right)+15 \Omega _m^3\nonumber\\+& \Omega _m \Omega _r \left(51 \Omega
_r-31\right)+16 \left(\Omega _r-1\right) \Omega _r^2\big)\bigg],
\end{align}
where prime denotes derivative with respect to $\ln a$ and	
\begin{align}
A=&6 \left(\Omega _m+\Omega _r\right){}^3 \left(2 \Omega _m+2 \Omega _r-1\right) \left(3
\Omega _m+4 \Omega _r\right)\nonumber\\&-\alpha _1 \left(\Omega _m+\Omega _r\right) \left(3
\Omega _m+4 \Omega _r\right){}^3.
\end{align}
The above dynamical system has three fixed points which correspond to matter dominated, radiation and de Sitter evolutions. The radiation dominated fixed point with $\omega_{eff}\equiv -1-2\dot{H}/3 H^2=1/3$ is
$$P_r=(\Omega_m ,\Omega_r)=\left(0,1+ \f83\alpha_1\right).$$
The eigenvalues of this fixed point are $-4$ and $1$. As a result, we have an saddle, unstable radiation fixed point.  We have also a matter dominated fixed point $\omega_{eff}=0$ as
$$P_m=(\Omega_m ,\Omega_r)=\left(1+ \f32\alpha_1,0\right),$$
with eigenvalues are $-1, -3$.‌This shows that the matter dominated fixed point is stable in this space-time. There is also an asymptotic de Sitter fixed point
$$P_\Lambda=(\Omega_m ,\Omega_r)=\left(0,0\right),$$
for which we have $\omega_{eff}=-1$. This fixed point is in fact non-hyperbolic and the standard linearization does not reflect the true behavior of the system. However, calculations reveal that this point is semi-stable and in addition the unstable manifold of the radiation fixed point $P_r$ can end at $P_\Lambda$. In figure \eqref{figdyna} we have plotted the phase portrait corresponding to the dynamical system \eqref{dy1} and \eqref{dy2}. One can see from the figure that for a perfectly radiation dominated initial configuration, the flow will end at the unstable fixed point $P_r$. For other initial values the universe will end at the stable matter dominated fixed point $P_m$. In the figure, we have plotted the unstable manifold of the fixed point $P_r$. One can see that there are two possible evolutions for the universe which is initially in a radiation dominated phase. For the first one, denoted by ``a", the universe will end at the matter dominated phase $P_m$. However, there is another possibility, denoted by ``b", in which the universe will end at the de Sitter phase. This later case is the one we have considered in this paper. 
\section{Energy Conditions} 
In this section, we are going to obtain the energy conditions associated with the theory \eqref{action}. Note that the metric field equation can be rewritten in the form
\begin{align}
\kappa^2G_{\mu\nu}=\f12 T^{eff}_{\mu\nu},
\end{align}
where we have defined the effective energy momentum tensor $T^{eff}_{\mu\nu}$ as
\begin{align}
T&^{eff}_{\mu\nu}=T_{\mu\nu}\nonumber\\&-2\beta\bigg(g_{\mu\nu}\nabla^\alpha f\nabla_\alpha\nabla_\beta f\nabla^\beta f+\Box f \nabla_\mu f \nabla_\nu f\nonumber\\&-2\nabla^\alpha f\nabla_\alpha\nabla_{(\mu}f\nabla_{\nu)}f \bigg)-\f{2\alpha\Lambda}{L_m}(T_{\mu\nu}-g_{\mu\nu}L_m)\Box f\nonumber\\&-\f{2\beta\Lambda}{L_m}(T_{\mu\nu}-g_{\mu\nu}L_m)\bigg((\Box f)^2-\nabla_\alpha\nabla_\beta f\nabla^\alpha\nabla^\beta f\nonumber\\&-R_{\alpha\beta}\nabla^\alpha f\nabla^\beta f\bigg)-2\alpha\left(\nabla_\mu f\nabla_\nu f-\f12g_{\mu\nu}\nabla_\alpha f\nabla^\alpha f \right).
\end{align}
In the cosmological background, one can obtain the effective energy density and pressure as
\begin{align}
\rho_{eff}=\rho+6 \beta\Lambda ^3  H  \Xi^3 -\alpha  \Lambda ^2 \Xi^2,
\end{align}
and
\begin{align}
p_{eff}&= \omega  \rho +6 \beta  \Lambda ^3 \omega  \Xi^2 \dot{H}+6 \beta  \Lambda ^3 \Xi^2 \dot{H}+12 \beta 
\Lambda ^3 \omega  H \Xi \dot{\Xi}\nonumber\\&+12 \beta  \Lambda ^3 H \Xi \dot{\Xi}-6 \alpha 
\Lambda ^2 \omega  H \Xi-6 \alpha  \Lambda ^2 H \Xi\nonumber\\&+18 \beta  \Lambda ^3
\omega  H^2 \Xi^2+18 \beta  \Lambda ^3 H^2 \Xi^2-2 \alpha  \Lambda ^2 \omega 
\dot{\Xi}\nonumber\\&-2 \alpha  \Lambda ^2 \dot{\Xi}-2 \beta  \Lambda ^3 \Xi^2 \dot{\Xi}-\alpha  \Lambda ^2\Xi^2,
\end{align}
where we have introduced $\Xi=\dot{\rho}/\rho$, and $H$ is the Hubble parameter. For the baryonic matter content of the universe, we have assumed the barotropic equation of state of the form $p=\omega\rho$ with $\omega\geq0$.

There are four main energy conditions well-known as the weak energy condition (WEC), strong energy condition (SEC), null energy condition (NEC) and dominant energy condition (DEC), which can be summarized for a perfect fluid as (see \cite{EC} for details)
\begin{align}
&\textmd{WEC}:\quad \rho_{eff}\geq0,\quad\rho_{eff}+p_{eff}\geq0,\nonumber\\
&\textmd{SEC}:\quad\rho_{eff}+p_{eff}\geq0,\quad\rho_{eff}+3p_{eff}\geq0,\nonumber\\
&\textmd{NEC}:\quad \rho_{eff}+p_{eff}\geq0,\nonumber\\
&\textmd{DEC}:\quad \rho_{eff}\geq0,\quad\rho_{eff}\pm p_{eff}\geq0.\\\nonumber
\end{align}
In an accelerated expanding universe, the SEC should be violated. This means that we should have $\rho_{eff}+3p_{eff}<0$. 

Now, consider the late-time cosmology in which the Hubble parameter varies slowly and one can assume that $H\approx H_0$ and $\dot{H}\approx0$. Also suppose that the energy density behaves like $\rho\propto a^{-n}$ with $n$‌ is an arbitrary positive number. In this case $\Xi$ also varies slowly and we have $\Xi\approx-nH_0$.‌ The conditions for which the SEC is violated and the rest remain satisfied can be written as the following set of inequalities
\begin{align}
-\alpha -6 n\beta  \Lambda H_0^2   \geq0,\nonumber\\
-(n-3 (\omega +1)) \left(\alpha +3 n \beta \Lambda  H_0^2 \right)\geq0,\nonumber\\
-\alpha  (\omega +1)-n(n+3 \omega +3)\beta \Lambda H_0^2\geq0,\nonumber\\
\alpha  (-2 n+9 \omega +9)-3 \beta     n (n-9 (\omega
+1))\Lambda H_0^2<0.
\end{align}
The above system can be solve to obtain
\begin{align}
\alpha\leq0,\qquad \beta <\frac{9 \alpha  \omega +9 \alpha -2 \alpha  n}{3n (n-9 \omega -9) 
	\Lambda  H_0^2},
\end{align}
where
$3 (\omega +1)\leq n<9 (\omega +1)$. Note that the lower bound for $n$ corresponds to the standard GR ‌behavior of the fluid. Assuming $\alpha=-1/2$‌ which gives the canonical kinetic term for $f$‌ in \eqref{action}, one can see that for dust dominated universe with $\omega=0$, we obtain $3\leq n<9$. This gives
\begin{align}
\beta\Lambda H_0^2<\frac{9-2 n}{54 n-6 n^2},
\end{align}
which gives $\beta\Lambda H_0^2<1/36$ for $n=3$, so the accelerated expansion also occurs in the case of vanishing $\beta$. Note that the estimated value for $\beta$ in the previous section is in agreement with the values of $\beta$ obtained here.

The same is true also for a radiation dominated universe with $\omega=1/3$ where $4\leq n<12$ and
\begin{align}
\beta \Lambda H_0^2<\frac{n-6}{3 (n-12) n}.
\end{align}
This gives $\beta\Lambda H_0^2<1/48$ for $n=4$. As a result one can see that the accelerated expansion of the universe can be achieved with baryonic matter at late times.
\section{Brans-Dicke correspondence}
The standard Einstein-Hilbert action could in principle produce a cosmological constant for a special matter dependent conformal transformation. Consider a conformal transformation of the form
$
g_{\mu\nu}\rightarrow \tilde{g}_{\mu\nu}=\Omega^2 g_{\mu\nu},
$
where $$\Omega=\f{\Lambda}{(2\lambda)^{1/4}}\exp{\f{f}{4\Lambda}},$$ and $\lambda$ is some constant. The Ricci scalar transforms as
\begin{align}
R=\Omega^2 \tilde{R} +\f{3}{2\Lambda}\Box f,
\end{align}
where $\tilde{R}$ is the Ricci scalar constructed from $\tilde{g}_{\mu\nu}$.
Under the above transformation, the Einstien-Hilbert action
$$S=\int d^4 x \sqrt{-g}(\kappa^2 R+\mc{L}_m),$$
reduces to
\begin{align}\label{new1}
S=\int & d^4 x \sqrt{-\tilde{g}}\bigg[\sqrt{\f{2\lambda}{|L_m|}}\kappa^2\tilde{R}
-2\lambda\bigg].
\end{align}
Here, one can consider $\lambda$ as an effective cosmological constant, which in principle could produce accelerated expansion. However, because there is a non-minimal coupling between the Ricci scalar and the inverse of the matter Lagrangian, the theory \eqref{new1} can not give an acceleration. This is due to the fact that the matter Lagrangian damps at late times and as a result compensate the effect of $\lambda$.
In our theory however, we have two other terms related to $\alpha$‌ and $\beta$ in \eqref{action} which makes the late time acceleration possible. The transformed forms of the $\alpha$ and $\beta$ terms can be obtained as
\begin{align}
\sqrt{-g} \nabla_\mu f \nabla^\mu f=\Omega^{-2}\sqrt{-\tilde{g}} \tilde{\nabla}_\mu f \tilde{\nabla}^\mu f,
\end{align}
\begin{align}
\sqrt{-g}\Box f\nabla_\mu f \nabla^\mu f=\sqrt{-\tilde{g}}\tilde{\nabla}_\mu f \tilde{\nabla}^\mu f\left(\tilde{\Box}f-\f{1}{2\Lambda}\tilde{\nabla}_\nu f \tilde{\nabla}^\nu f\right),
\end{align}
where $\tilde{\nabla}$ is the covariant derivative made by $\tilde{g}_{\mu\nu}$. The theory \eqref{action} is then transformed as
\begin{align}\label{jor}
S=\int & d^4 x \sqrt{-\tilde{g}}\bigg[\Omega^{-2}\left(\tilde{R}+\alpha\tilde{\nabla}_\mu f \tilde{\nabla}^\mu f \right)\nonumber\\&
+\beta \tilde{\nabla}_\mu f \tilde{\nabla}^\mu f\left(\tilde{\Box}f-\f{1}{2\Lambda} \tilde{\nabla}_\nu f \tilde{\nabla}^\nu f\right)
-2\lambda\bigg],
\end{align}
with a manifest cosmological constant term.
\section{conclusions}
In this paper, we have considered a theory containing a minimal derivative matter coupling with geometry which possesses a dS expansion at late times. The matter interactions is expressed in terms of a new function $f$ which is defined as a Logarithm of the matter Lagrangian $L_m$. We have added a cubic Galileon like interaction term in the action for completeness. Also, we have shown that the late time dS expansion can be achieved without this term. However, the main result of the paper is that the dS expansion can be achieved from the baryonic matter content of the universe without introducing dark sectors. From the cosmological solutions, one can see that the dust energy density dominates from the radiation energy density at late times as is expected observationally.

The dynamical system analysis of the theory shows an unstable radiation dominated hyperbolic fixed point which its unstable manifold can end at an asymptotic semi-stable dS fixed point. There is also a stable hyperbolic matter dominated fixed point in the theory. For the vast range of initial conditions, the dynamical trajectories of the theory approach the radiation dominated fixed point and then ends at the stable matter dominated fixed point. For purely radiation dominated initial conditions, the trajectories can end at the dS fixed point. The dynamical behavior of the theory will alter if we consider higher order Galileon/Horndeski interactions, which will be the scope of our future work. 

It should be noted that the matter energy momentum tensor is not conserved in general but in the cosmological setup with a perfect fluid matter, it is conserved. The theory has also a well-behaved Newtonian limit with an effective cosmological constant.

 It should be noted that one can use the trace of the energy momentum tensor $T$ instead of the Lagrangian density $L_m$ in the definition of $f$. For a perfect fluid, one has $T=-\rho+3p$. As a result one expects that the qualitative behavior of using $T$ is the same as ours (using $L_m$), possibly with some redefinition of the constant $\Lambda$.

\begin{acknowledgments}
	The authors would like to thank Mohammad Reze Razvan and Sheida Shahidi for very useful discussions on dynamical analysis of the model.
\end{acknowledgments}


\begin{thebibliography}{99}
\bibitem{Einstein} A. Einstein, Sitzungsberichte der Königlich Preussischen Akademie der Wis-
senschaften (Berlin), 1919, p. 349.
\bibitem{PDU} Z. Haghani, T. Harko and S. Shahidi, Phys. of the Dark Univ. 21 (2018) 27.
\bibitem{rastall} P. Rastall, Phys. Rev. D 6 (1972) 3357.
\bibitem{fRT} T. Harko, F.S.N. Lobo, S. Nojiri, S.D. Odintsov, Phys. Rev. D 84 (2011) 024020.
\bibitem{Lmtheory} O. Bertolami, C.G. Boehmer, T. Harko, F.S.N. Lobo, Phys. Rev. D 75 (2007)
104016; 
\bibitem{ourRT} Z. Haghani, T. Harko, F.S.N. Lobo, H.R. Sepangi, S. Shahidi, Phys. Rev. D 88
(2013) 044023.
\bibitem{roshan}M. Roshan and F. Shojai, Phys. Rev. D 94 (2016) 044002; , J. D. Barrow and C. Board, Phys. Rev. D 96 (2017) 123517.
\bibitem{powered} O. Akarsu, N. Katırcı and S. Kumar, Phys. Rev. D 97 (2018) 024011.
\bibitem{EC}  S. W. Hawking and G. F. R. Ellis, The Large Scale Structure of Space-Time (Cambridge University Press, UK, 1973).

\end{thebibliography}
\end{document}